\documentclass[prc,twocolumn,showpacs,amsmath,nofootinbib,amssymb]{revtex4-1}
\usepackage{dcolumn}
\usepackage{bm}
\usepackage{color}
\usepackage{amssymb}
\usepackage{amsmath}
\usepackage{graphicx}
\usepackage{amsfonts}
\usepackage{slashed}
\usepackage{pstricks}
\usepackage{float}
\usepackage{hyperref}
\usepackage{array}
\allowdisplaybreaks
\usepackage{dcolumn}
\usepackage{epsf}
\usepackage{isotope}
\begin{document}
\title{Comment on ``Quasielastic lepton scattering and back-to-back nucleons \\ in the short-time approximation'', by S. Pastore {\em et al.}}
\author{Omar Benhar}
\email{omar.benhar@roma1.infn.it}
\affiliation{INFN and Department of Physics,~Sapienza University, I-00185 Rome, Italy}

\date{\today}

\begin{abstract}
The article of Pastore {\em et al.}, while proposing an interesting and  potentially useful 
approach for the generalisation of Quantum Monte Carlo techniques to the treatment of the nuclear electromagnetic  response, 
features an incorrect and misleading discussion of $y$-scaling. 
The  response to interactions with transversely polarised virtual photons receives sizeable contributions from 
non-scaling processes, 
in which the momentum transfer is shared between two nucleons. It follows that, contrary to what is stated by the the authors, 
$y$-scaling in the transverse channel is accidental. 

\end{abstract} 

\index{}\maketitle

The work of Pastore {\em et al.}~\cite{pastore} can be seen as a first step towards the implementation of the 
factorisation scheme, which naturally emerges from the formalism of the impulse approximation~\cite{noemi:PRL}, in the computational framework of Quantum 
Monte Carlo (QMC). 
In view of the difficulties associated with the identification of specific final states in 
the nuclear responses obtained from QMC calculations, this is an interesting, and potentially useful, development.

The Short Time Approximation (STA) developed by the authors involves a number of strong simplifying assumptions\textemdash such 
as neglect of the energy dependence in the propagator of the spectator system, analysed in Ref.\cite{ambiguity}\textemdash the validity 
of which will only be fully appraised in years to come, 
when the proposed approach will be extended to a broader kinematical range and  
nuclear targets other than the three- and four-nucleon systems.
The discussion of scaling in Section IV, on the other hand, comprises incorrect and misleading statements requiring a prompt clarification. 

The occurrence of $y$-scaling in electron-nucleus scattering\textemdash that is, the observation that the target response, which in general depends
on both momentum and energy transfer, ${\bf q}$ and $\omega$, in the limit of 
large $q = |{\bf q}|$ can be reduced to a function of the single variable $y=y(q,\omega)$ ~\cite{yscaling:3He,yscaling:C}\textemdash
reflects the onset of the kinematical regime in which the dominant reaction mechanism is elastic scattering off individual nucleons~\cite{west}.
 
The definition of the scaling variable $y$ follows from the assumption that the momentum transfer is absorbed by {\em only one}
 nucleon. In the absence of Final State Interactions (FSI) between the struck nucleon and the spectators, 
conservation of energy in the lab frame entails the relation~\cite{CPS1}
\begin{align}
\label{def:y}
\omega + M_A  & = \sqrt{ m^2 + (y + q)^2}\\
\nonumber
& + \sqrt{ (M_A -m + E_{\rm thr})^2 + y^2 } \ ,
\end{align}
where $M_A$ and $m$ are the target and nucleon mass, respectively,  while $E_{\rm thr}$ denotes the nucleon emission threshold.
Equation~\eqref{def:y} shows that the scaling variable has a straightforward physical interpretation, being trivially related 
to the projection of the momentum of the struck nucleon along the direction  of  the momentum transfer, $k_\parallel = {\bf k} \cdot {\bf q}/q$.

The scaling function of a nucleus of mass number ${\rm A}$ and charge ${\rm Z}$, defined as~\cite{CPS1}
\begin{align}
F(y) = \lim_{q \to \infty} F(q,  y)
\end{align}
is obtained from
\begin{align}
F(q , y) = \frac{ d \sigma_{eA} } {{\rm Z} d \sigma_{ep} + ({\rm A}-{\rm Z}) d \sigma_{en} } \Big( \frac{ d \omega} { d k_\parallel } \Big) \ , 
\end{align}
where $d \sigma_{eA}$ is the measured nuclear cross section, while $d \sigma_{ep}$ and $d \sigma_{en}$ are the elastic electron-proton
and electron-neutron cross sections, stripped of the energy-conserving $\delta$-function\footnote{Here, the elementary electron-nucleon cross sections, which explicitly depend on the nucleon momentum, ${\bf k}$, and and removal energy, $E$, are evaluated at  $|{\bf k}| = |y|$ and $E=E_{\rm  thr}$.}.
Large deviations from the scaling behaviour, observed at $y>0$, arise from processes other than elastic scattering, while 
smaller scaling violations at $y<0$ are ascribed to FSI~\cite{CPS1}. 

The above definitions imply that the scaling function is an intrinsic property of the target, providing information on the nucleon momentum distribution, $n({\bf k})$. In deuteron, the relation between scaling function and momentum distribution takes the simple  form~\cite{CPS2}
\begin{align}
n(k) = -\frac{1}{2\pi} \frac{1}{y} \left. \frac{dF(y)}{dy} \right|_{|y| = k} \ ,
\end{align}
with $k = |{\bf k}|$.

In the 1980s, Finn {\em et al.}~\cite{scaling:LT} performed the first scaling analysis of the carbon responses to interactions with longitudinally 
and transversely polarised virtual photons, measured at Saclay~\cite{barreau}. The results of this work 
revealed a significant excess of strength in the transverse channel, which, however, did not appear to spoil the scaling behaviour at $y<0$.
As a consequence, the analysis led to the determination of two distinct $q$-independent functions, $F_L(y)$ and $F_T(y)$, even though
the interpretation of $F_T(y)$ as a scaling function cannot be reconciled  with the presence of contributions arising from non-scaling processes, driven by two-nucleon currents.  
More recently, similar results have been obtained from the analysis of the longitudinal and transverse responses of light nuclei~\cite{scaling:LT2}.

Pastore {\em et al.} fail to introduce the reader to the concept of scaling and the interpretation of the scaling variable. They limit themselves to assert that, because the results of their Green's Function Monte Carlo (GFMC) calculations reproduce the experimental data, they {\em obviously scale}.
Regretfully, this statement is meaningless and misleading\footnote{The reader is, in fact, also misled into believing that Fig.~3 of the paper of Carlson 
{\em et al.}~\cite{scaling:LT2} show longitudinal and transverse response functions, while the quantities displayed are actually the functions
$F_L(y)$ and $F_T(y)$ defined by Eqs.(27)-(29) of Ref.~\cite{def:fy}.}

In Section IV, 
the authors go to considerable lengths to explain the mechanism leading to $q$-independence of the GFMC 
response functions. However, their conclusion that $y$-scaling is preserved even in the presence of a mechanism 
other that single-nucleon knock out 
does not take into account the fact that $q$-independence and $y$-scaling are 
distinct properties, and do not necessarily imply one another.

Accidental $y$-scaling\textemdash that is, scaling in the presence of non-scaling mechanisms, such as FSI, giving rise to sizeable $q$-independent contributions to the nuclear response\textemdash is long known to occur in a variety 
of processes, ranging from electron-nucleus scattering~\cite{accidental} to neutron scattering from liquid helium~\cite{negele}.
Obviously, when scaling is accidental, the interpretation of both the scaling variable and the scaling function discussed above
is no longer applicable. 

Processes involving two nucleon currents {\em do not scale} in the variable $y$, because the momentum transfer is shared between two nucleons, 
and conservation of energy cannot be written as in Eq.~\eqref{def:y}. It follows that $y$-scaling in the transverse channel is, in fact, accidental.  
As correctly noted by the authors of Ref.~\cite{scaling:LT}, a meaningful scaling function, providing information on initial-state dynamics, 
can only be obtained from the analysis of the longitudinal response. 

On the constructive side, it should be noted that, after removal of the excess transverse strength arising form processes involving two-nucleon currents, the longitudinal and transverse responses 
obtained by Pastore~{\em et~al} may be employed to perform a fully consistent study of the universality of the scaling function. 
The results of such a study would be valuable for the ongoing efforts to exploit $y$-scaling as a tool 
for the analysis of 
 the signals detected by accelerator-based searches of neutrino oscillations~\cite{superscaling}. \\
 

\end{document}